%% file: paper.tex
\pdfoutput=1
\documentclass[pdflatex,a4paper]{article}

\usepackage{color,graphicx}
\usepackage{amsmath,amsthm}
\usepackage{newtxtext}
\usepackage[varg]{newtxmath}
\usepackage{cite,url}
\usepackage{xspace} 
\usepackage{authblk} 
\usepackage{multirow}

\usepackage[
 pdftex,
 colorlinks=true,
]{hyperref}

\input{./symdef.tex}

\newcommand\schemeabbrv{$\mu$ODNS\xspace}
\hypersetup{
  pdfauthor={Jun Kurihara and Takeshi Kubo},
  pdftitle={Mutualized oblivious DNS (\schemeabbrv): Hiding a tree in the wild forest}
}
\title{Mutualized oblivious DNS (\schemeabbrv):\\ Hiding a tree in the wild forest}
\author[1,2,3]{Jun Kurihara}
\author[2]{Takeshi Kubo}
\affil[1]{Graduate School of Information Science, University of Hyogo}
\affil[2]{Zettant Inc.}
\affil[3]{Advanced Telecommunications Research Institute International}
\affil[ ]{{\rm kurihara@ieee.org}}
\begin{document}

\maketitle

\begin{abstract}
The traditional Domain Name System (DNS) lacks fundamental features of security and privacy in its design.
As concerns of privacy increased on the Internet, security and privacy enhancements of DNS have been actively investigated and deployed.
Specially for user's privacy in DNS queries, several relay-based anonymization schemes have been recently introduced, however, they are vulnerable to the collusion of a relay with a full-service resolver, i.e., identities of users cannot be hidden to the resolver.
This paper introduces a new concept of a multiple-relay-based DNS for user anonymity in DNS queries, called the \emph{mutualized oblivious DNS} (\schemeabbrv), by extending the concept of existing relay-based schemes.
The \schemeabbrv introduces a small and reasonable assumption that each user has at least one trusted/dedicated relay in a network and mutually shares the dedicated one with others.
The user just sets the dedicated one as his \emph{next-hop}, first relay, conveying his queries to the resolver, and randomly chooses its \emph{$0$ or more} subsequent relays shared by other entities.
Under this small assumption, the user's identity is concealed to a target resolver in the \schemeabbrv even if a certain (unknown) subset of relays collude with the resolver.
That is, in \schemeabbrv, users can preserve their privacy and anonymity just by paying a small cost of sharing its resource.
Moreover, we present a PoC implementation of \schemeabbrv that is publicly available on the Internet.
We also show that by measurement of round-trip-time for queries, and our PoC implementation of \schemeabbrv achieves the performance comparable to existing relay-based schemes.
\end{abstract}

\section{Introduction}

%

\emph{Domain Name System} (DNS) plays a role to map human-readable host names to machine-readable information on the Internet; a user (or a \emph{stub resolver}) exchanges a host name with its IP address and associated resource records by querying a \emph{full-service resolver}.
In the traditional DNS called Do53, this exchange of DNS messages, i.e., a user's \emph{query} and a resolver's \emph{response}, is performed over UDP or TCP on port 53 in the form of plaintext.
This means that in Do53, the user's Internet activity could be easily exposed to monitoring authorities.
Recent increased demands for privacy on the Internet have been motivating to introduce new extensions for the privacy enhancement of DNS.

From the above demands, there have been proposed some \emph{encryption} schemes for DNS messages \cite{dnscurve,dnscrypt,dot-rfc,doh-rfc} to avoid them from being wiretapped.
In these schemes, a secure encryption channel is established by a certain public key cryptography between a user and an encryption-enabled full-service resolver, and DNS messages are exchanged via the channel.
Hence users' activities in DNS can be protected by these schemes from censorship authorities monitoring messages transported.
However, another privacy concern has been raised here even if such encryption techniques are employed:
While DNS messages are encrypted, full-service resolvers see plaintext messages from/to users by the nature of DNS, and they can fully observe any activities of users in DNS.

In order to resolve this second privacy concern, there have been proposed several \emph{anonymization} techniques \cite{Schmitt2019,Singanamalla2020,Denis2021,Denis2020} to hide a user's IP address from a full-service resolver.
\emph{Oblivious DNS} (ODNS) \cite{Schmitt2019} is such the first anonymization scheme, which was designed to be compatible with the standard DNS architecture.
In the scheme, an ODNS-specialized resolver is introduced, and an existing full-service resolver over Do53 is leveraged as a \emph{relay} forwarding encrypted DNS messages between the ODNS resolver and a user.
Thus the user's address is hidden from the ODNS resolver thanks to the relay.
By adopting this relay-based concept into encrypted DNS schemes \cite{doh-rfc,dnscrypt}, \emph{Anonymized DNSCrypt} (ADNSCrypt) \cite{Denis2021,Denis2020} and \emph{Oblivious DNS over HTTPS} (ODoH) \cite{Singanamalla2020} have recently been introduced.
Unlike ODNS, these schemes directly introduce intermediate relays dedicated to the schemes between users and encryption-enabled resolvers and omit the compatibility with Do53.
This simplification in the architecture results in their good performance comparable to standard encrypted DNS schemes \cite{Singanamalla2020}.

One main drawback of the above relay-based anonymization schemes is the lack of \emph{collusion resistance}: the privacy could be completely corrupted when the relay colludes with the target resolver.
Considering the current deployment of DNS, users do not have various choices of full-service resolvers enabling DNS message encryption, and usually use ones operated by large and limited entities \cite{Deccio2019}, e.g., Google, Cloudflare, Quad9, etc.
Much like encryption-enabled resolvers, relays would be operated by such limited big players as mentioned in \cite[Section 7.1]{Singanamalla2020}.
Hence it may increase concerns about collusion and surveillance.
If a user cannot trust such existing relays, the simplest solution should be to build a \emph{dedicated} relay to the user.
However, this is completely useless since the relay's address exposed to the target resolver can be uniquely associated with the user itself.
In other words, every user must \emph{unconditionally} trust and choose a public and shared relay relaying its message when we employ these anonymization schemes.
We thus have no way to fundamentally remove the concern on the collusion in these schemes.

This paper aims to solve the above problems of collusion in existing anonymization schemes, and to provide a practical scheme guaranteeing the anonymity of users under the condition that \emph{users cannot trust most of network nodes} for DNS.
To these ends, this paper introduces a new architectural concept to preserve the user's anonymity in DNS messages, which is called the \emph{mutualized oblivious DNS} (\schemeabbrv).
The \schemeabbrv is designed in such a way that by introducing a small and reasonable assumption on nodes adjacent to a user, it maintains moderate performance and preserves the anonymity in DNS against untrusted resolvers and network nodes that might be colluded.
From the perspective of its architecture, it could be regarded as an extension of existing relay-based schemes to the one allowing multiple relays and route randomization.
Thus this could be viewed as an approach similar to Tor \cite{tor} but is highly specialized and simplified to DNS.
Under the concept of \schemeabbrv, \emph{every user can preserve the anonymity of DNS messages just by paying a small cost even if some network nodes have colluded with the resolver}.
Furthermore, it is also guaranteed that every user can NOT corrupt the anonymity of any other users.
Our contributions in this paper are not only to introduce the concept of \schemeabbrv but also to present a proof-of-concept (PoC) implementation of \schemeabbrv\footnote{Available at \cite{modns-server,modns-proxy,modns-resolvers-relays}} based on ADNSCrypt \cite{Denis2021,Denis2020}.
This paper evaluates the \schemeabbrv by this PoC implementation deployed on the Internet and demonstrates that it would achieve comparable performance to ADNSCrypt and ODoH.
We thus claim that our anonymization concept is practical and reasonable from both perspectives of its design and performance in the severe environment for user privacy.

The remainder of this paper is organized as follows:
Section~\ref{sect:related_work} introduces the background of our work and summarizes recent studies on privacy in DNS.
Section~\ref{sect:formulation} presents assumptions and formally introduces problems considered in this paper.
Under the given assumptions, Section~\ref{sect:overview} overviews our concept of \schemeabbrv and explains how it works possibly in the presence of colluded nodes.
Section~\ref{sect:poc} explains our proof-of-concept implementation of \schemeabbrv and introduces its deployment on the Internet.
Section~\ref{sect:discussion} gives evaluations and discussions on the security, privacy, limitations and performance of \schemeabbrv and its PoC implementation.
Finally Section~\ref{sect:conclusion} concludes this paper.

\section{Background and related work}\label{sect:related_work}

Domain Name System (DNS) \cite{rfc1034,rfc1035} was originally standardized at IETF as a system that maps human-readable host names to machine-readable resource records such as IP addresses.
Since the traditional DNS has architectural problems causing security and privacy issues, several enhancements of DNS protocols have been investigated and incrementally deployed on the Internet.
For instance, there is no guarantee of the authenticity of resource records in the vanilla DNS,
and hence \emph{DNS Security Extension} (DNSSEC), e.g., \cite{rfc4033,rfc4034,rfc4035}, has been proposed to protect users from attacks like DNS cache poisoning.

Recently, as increasing demands for privacy on the Internet, e.g., \cite{rfc7258}, another architectural problem of DNS has been drawing attention, which is the lack of \emph{confidentiality} and \emph{anonymity} of exchanged messages.
In the following, we shall summarize recent efforts for the problem.

\subsection{Encrypted DNS protocols}
In the traditional DNS via TCP/UDP port 53 (Do53), query messages and response messages are exchanged between a user and a full-service resolver in \emph{plaintext}.
This implies that with Do53, the user's activity on the Internet can be easily exposed to eavesdroppers on a channel, and also that DNS messages are immediately censored by a certain authority observing channel.
In order to protect DNS messages from being eavesdropped, several \emph{encrypted DNS} protocols have been investigated \cite{dnscurve,dnscrypt,doh-rfc,dot-rfc}.

In \emph{DNSCurve} \cite{dnscurve} and its successor \emph{DNSCrypt} \cite{dnscrypt}, query messages and response messages are encrypted with public keys exchanged between a user and a full-service resolver, and they are simply transported as UDP or TCP packets.
On the other hand, following the public key infrastructure (PKI),
\emph{DNS over TLS} (DoT) \cite{dot-rfc} establishes a transport layer security (TLS) connection between a user and a full-service resolver,
 and they securely exchange query and response messages over the connection.
\emph{DNS over HTTPS} (DoH) \cite{doh-rfc} leverages an HTTPS connection as the underlying secure layer, and the message exchange in DoH is executed through POST or GET methods.
We note that in these protocols, query and response messages are directly exchanged between a user (a stub resolver) and a full-service resolver as well as Do53.
In general, users need to utilize \emph{public resolvers}, e.g., Google, Cloudflare, Quad9, instead of ISPs' resolvers to enable these encrypted DNS protocols \cite{Deccio2019}.

\subsection{Anonymized/Oblivious DNS protocols}
The use of encrypted DNS protocols helps users to protect their privacy from being exposed to eavesdroppers on a channel.
However, because of the fundamental mechanism of DNS, plaintext query messages of users must be observed at full-service resolvers to search associated resource records.
This means that every plaintext query can be uniquely coupled with the user's identity, e.g., IP address, at the target resolver.
Hence several anonymization techniques for DNS queries have been recently investigated as enhancements of encrypted DNS, i.e., as an extra layer for anonymization, to decouple the user's identity from queries observed at the resolver \cite{Schmitt2019,Denis2021,Denis2020}.

First of all, the employment of Tor \cite{tor} is a straightforward technique for the anonymization of DNS queries.
Much like Web services, we can anonymize DNS queries by using the TCP-based\footnote{Tor supports only TCP.} encrypted DNS protocol over Tor, e.g., \emph{DNS over HTTPS over Tor} (DoHoT).
This approach usually involves a significant performance loss due to various reasons, e.g., the overhead of multiple-layered encryption, a large volume of traffic at Tor nodes, large RTT among geographically distributed nodes, etc.
Unlike this generic approach based on Tor, there have been proposed several anonymization protocols dedicated to DNS as follows.

\emph{Oblivious DNS} (ODNS) \cite{Schmitt2019} is the first protocol specialized for DNS anonymization.
In ODNS, an ODNS resolver is introduced and works as an authoritative name server of a special top level domain ``\texttt{.odns}''.
Each user generates a query to the \texttt{.odns} domain containing an encrypted original query.
An ODNS resolver decrypts the received encrypted query and dispatches the plaintext version of the query upstream on behalf of the user.
Then, a full-service resolver between the user and the ODNS resolver can be viewed as a \emph{relay} concealing the user's IP address from the ODNS resolver observing plaintext queries.

Although ODNS took an elegant approach compatible with the standard Do53, it involves a large round-trip-time (RTT) since all queries must be forwarded to the ODNS resolver, i.e., an authoritative server.
\emph{Oblivious DNS over HTTPS} (ODoH) \cite{Singanamalla2020,doh-ietf} has been designed by simplifying the architecture and omitting the compatibility with Do53, and realizes a performance comparable to the standard DoH.
ODoH introduces a relay called \emph{oblivious proxy} that just relays encrypted queries and responses between a user and a target resolver.
In ODoH, the target resolver is allowed to be separated into two nodes: a (Do53) full-service resolver and an encryption/decryption terminal called \emph{oblivious target}, where the oblivious target sends decrypted queries to the resolver on behalf of the user.
\emph{Anonymized DNSCrypt} (ADNSCrypt) \cite{Denis2021,Denis2020} has been designed from the concept fundamentally same as the ODoH from the viewpoint of the architecture, i.e., leveraging a relay node to hide the user's identity from the target resolver.
The architectural difference between ADNSCrypt and ODoH is only the existence of channel encryption based on TLS (HTTPS).
We can see that an ODoH relay, i.e., oblivious proxy, simply forwards encrypted DNS messages upstream/downstream much like an HTTPS proxy.
On the other hand, encrypted messages in ADNSCrypt are forwarded directly over UDP or TCP.

In ODoH and ADNSCrypt, each user chooses and fixes a single relay and every DNS message from/to the user goes through the relay.
Thus the relay hides the user's address from the target resolver's observation, and ODoH and ADNSCrypt decouple the user's IP address from DNS messages, i.e., queries.
As mentioned in \cite{Singanamalla2020}, this anonymity of the user is guaranteed unless the chosen relay colludes with the public resolver.
Additionally note that in these anonymized/oblivious DNS protocols, public resolvers like Google and Cloudflare, are assumed as target resolvers as well as the standard encrypted DNS.
Also note that the following assumption is implicitly posed in ODoH and ADNSCrypt: \emph{the chosen relay is used by multiple (ideally large number of) users. This is because the relay's identity like IP address must not be uniquely coupled with the user's as well.}

\section{Assumption and problem formulation}\label{sect:formulation}

\begin{figure}[tb]
\centering
\includegraphics[width=0.8\linewidth]{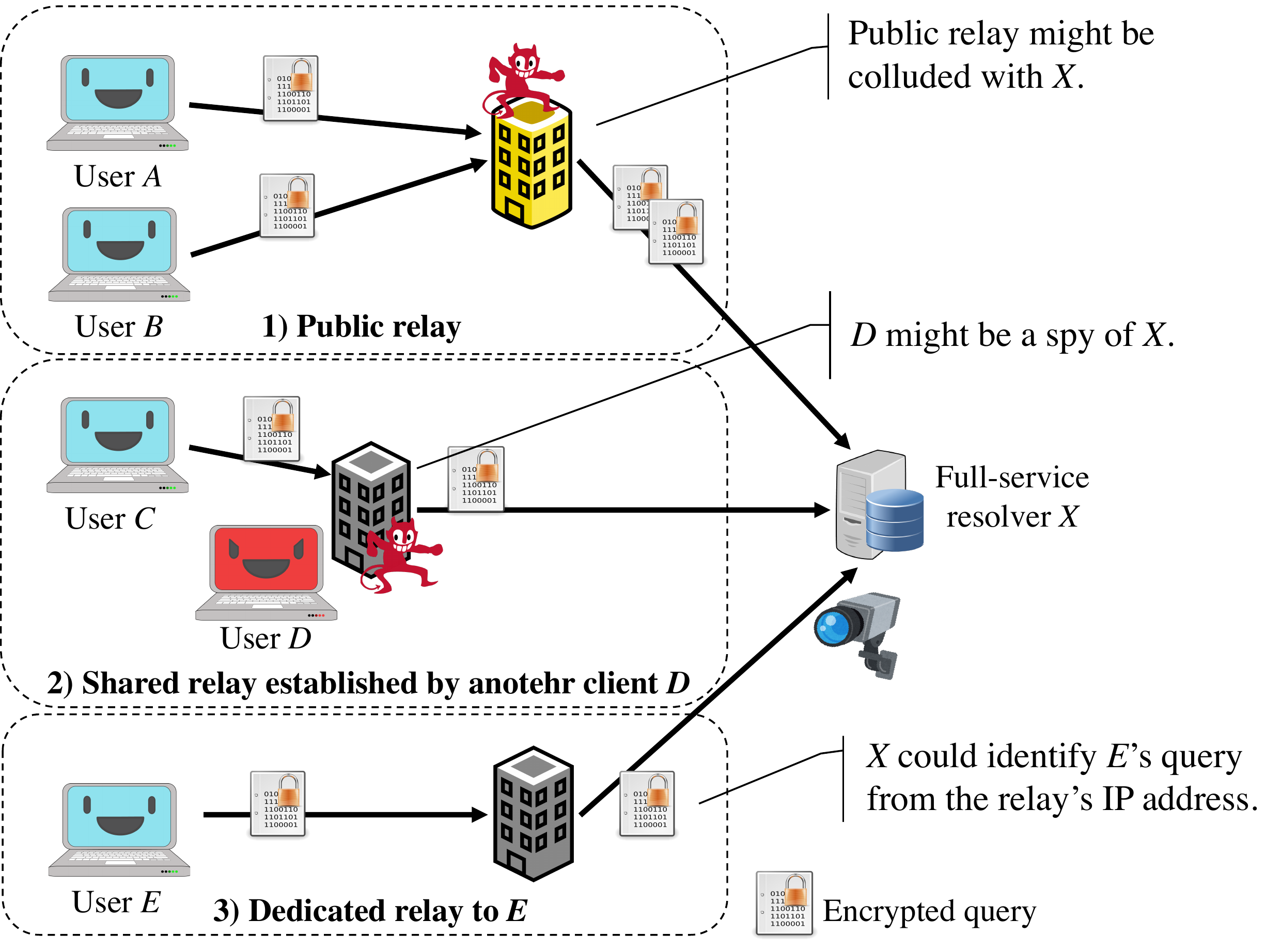}
\caption{Illustration of problems in existing relay-based schemes for user anonymity}
\label{fig:problem_of_proxy}
\end{figure}

This section formally presents our assumption and formulate the problem of the anonymity of DNS messages.

Our assumption is almost the same as that in ODoH \cite{Singanamalla2020} and ADNSCrypt \cite{Denis2020,Denis2021}.
First assume that a user will use a \emph{public resolver} as a full-service resolver such as Google DNS and Quad9, and that multiple (ideally large number of) users join the network.
Also, assume that there are network nodes called \emph{relays} that simply forward incoming encrypted DNS messages upstream or downstream.
As in ODoH and ADNSCrypt, relays are \emph{semi-honest}, i.e., they work correctly but they may try to observe the content of messages.
Hence, to avoid relays from observing messages, the user exchanges DNS messages with the resolver encrypted in an end-to-end manner.
This paper mainly focuses on attackers observing messages at public resolvers and relays, and we do not consider \emph{wiretappers} who observe transit messages on channels.

In addition to the above assumptions, we suppose that for a user, \emph{a relay operated by another entity can collude with the public resolver}, and the location of the colluded relay(s) is unknown to the user.
We realistically assume that a tiny subset of relays potentially colludes with the target resolver.
Other than potentially-colluded relays, \emph{every user (or every organization of a set of users) has at least one trusted instance of relays}, e.g., a relay deployed by himself in the network.
Unlike ODoH and ADNSCrypt, we do NOT assume public relays are unconditionally trusted by numbers of users, e.g.., a relay trusted by a user is NOT usually trusted by others.
In this environment, each user aims to exchange encrypted DNS messages with a public resolver in such a way that the resolver does not identify the user from received messages and information given by colluded relays.
This extra assumption can be viewed as the one relaxed from that in ODoH and ADNSCrypt from the viewpoint of users' trust on relays.
We see that this relaxation raises new problems in the usage of relays for users' anonymity in existing relay-based schemes, summarized in Figure~\ref{fig:problem_of_proxy}.

Here we enumerate such new problems when we employ the existing relay-based schemes.
Firstly, considering a \emph{public} relay expected to be used by multiple users, it is NOT always a trusted one for users by the assumption.
We face this concern as well when we choose a \emph{shared} relay operated by another user.
Namely, choosing a relay operated by another entity, we cannot remove a concern on being spied on because the chosen relay potentially might be colluded with the full-service resolver.
Secondly, although the only way to remove the concern is to choose and dedicate a relay trusted for each user, this approach is completely useless.
This is because the identity of the chosen relay can be uniquely bound with the identity of the user for the public resolver.
From these observations, \emph{we aim to design a scheme to utilize a trusted and dedicated relay of each user in such a way that the identity of the user is practically hidden from the target resolver} in this severe environment.
In the design of such a scheme, we also aim to avoid performance degradation and achieve moderate performance comparable to ODoH and ADNSCrypt.

\begin{remark}
We note that the above new problems of relay-based schemes might not occur in DoHoT due to the nature of Tor.
However, it involves incredible performance degradation as we stated in Section~\ref{sect:related_work}, and it is NOT a realistic approach for the requirement for real-time processing in DNS-like systems.
Indeed DNS-specific relay-based schemes have significantly outperformed a Tor-based scheme.
Hence we shall enhance the DNS-specific schemes to meet the performance requirement.
\end{remark}

To close this section, we give some justification for the assumption given above.
We claim that the additional assumption could be realistic: Considering deployment scenarios of ODoH relays discussed in \cite{Singanamalla2020}, relays are expected to be deployed by operators of existing public resolvers.
The user's anonymity in existing schemes relies only on third parties, and hence the user must \emph{unconditionally} trust them.
We remark that, however, this unconditional trust might not be always held for big players.

\section{Overview of \schemeabbrv}\label{sect:overview}
\begin{figure}[tb]
\centering
\includegraphics[width=\linewidth]{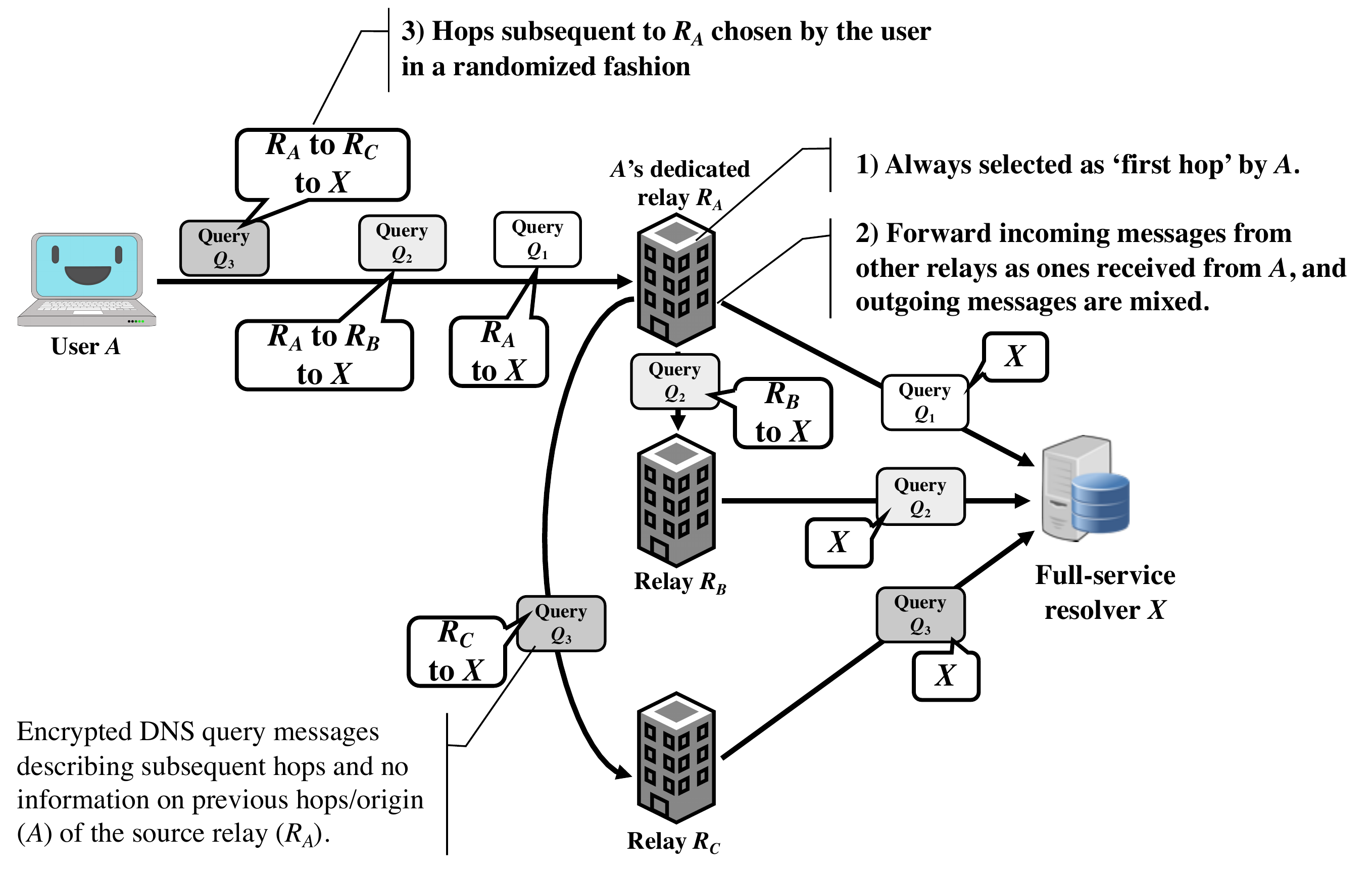}
\caption{Key features of \schemeabbrv}
\label{fig:modns_concept1}
\end{figure}
\begin{figure}[tb]
\centering
\includegraphics[width=\linewidth]{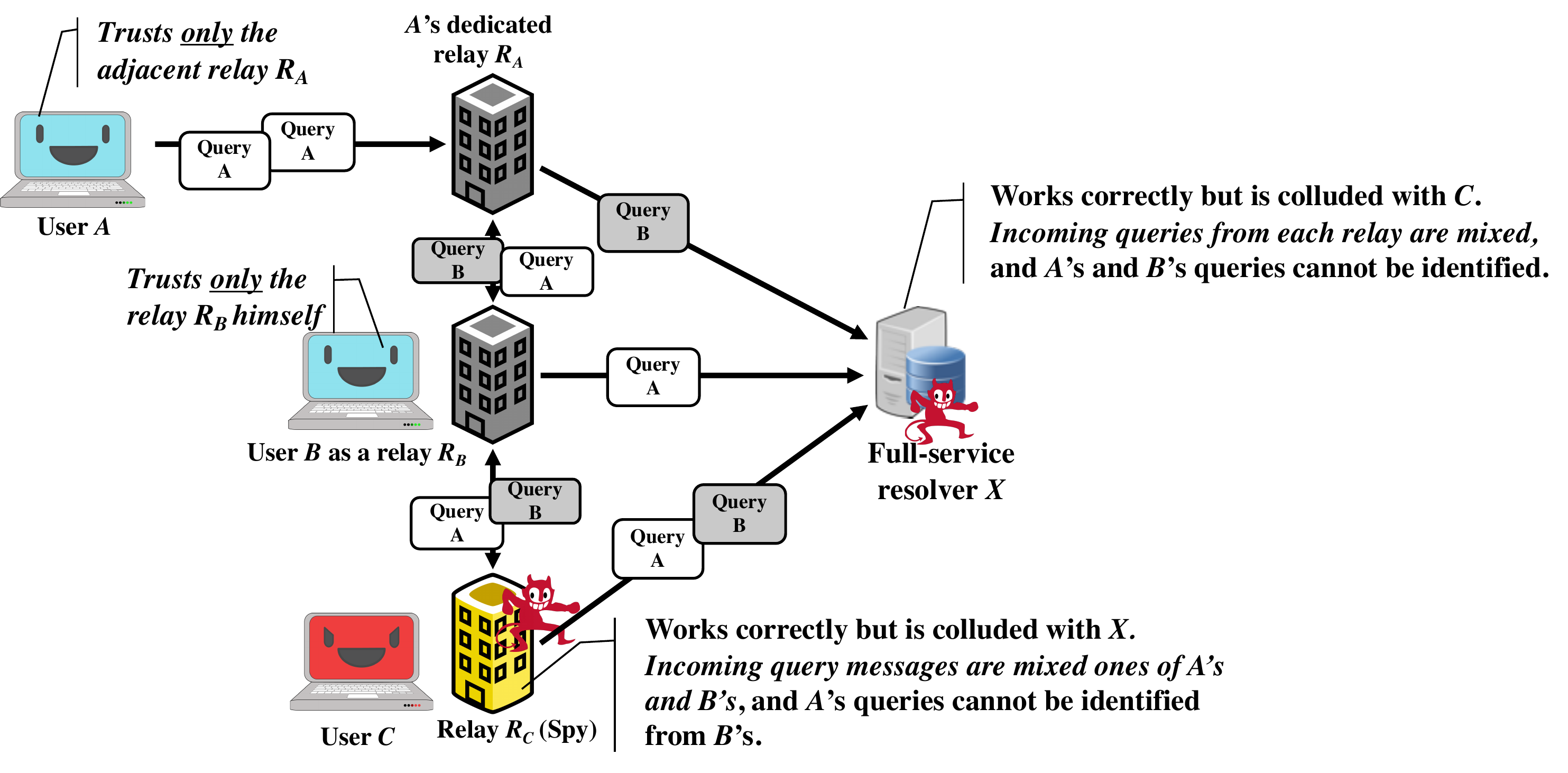}
\caption{An exemplary scenario of \schemeabbrv (3 parties case.), where user $A$ issues DNS query messages (query A's) routed as $R_A \rightarrow R_B \rightarrow X$ and $R_A \rightarrow R_B \rightarrow R_C$, and user $B$ issues ones (query B's) routed as $R_A \rightarrow X$ and $R_C \rightarrow X$. The resolver $X$ cannot see that the origin of messages received from $R_A$ is $B$, and also cannot identify origins of messages from $R_C$ even by utilizing the colluded spy $R_C$.}
\label{fig:modns_concept2}
\end{figure}

In the previous section, we supposed a severe environment: an unknown subset of relays have colluded with public resolvers and only his dedicated relay can be trusted.
The design goal of our scheme, \schemeabbrv, is to practically anonymize each user's DNS message in the environment.
Specifically, our goal is to establish a scheme that fully and mutually relies on each user's dedicated relay for collusion avoidance without leaking the user's identity from conveyed DNS messages.
To achieve the goal, we designed \schemeabbrv as a relay-based scheme leveraging \emph{multiple hops of relays} with several key features summarized as Figure~\ref{fig:modns_concept1}.
The key features of \schemeabbrv are enumerated as follows:
1) Employment of each user's dedicated relay(s) as his \emph{next-hop}.
2) Mutually sharing the dedicated relay to forward any incoming relayed DNS messages to other relays or the target resolver.
3) Randomization of user's choices for hops after his dedicated relay.
Figure~\ref{fig:modns_concept2} also illustrates an exemplary scenario achieved by the \schemeabbrv with these features.
In the following, we shall briefly describe these features.

\vspace{1ex}
\noindent\textsc{\textbf{1) Employment of a user's dedicated relay(s) as his next-hop:}}
Basically in the scheme, every DNS query message issued by a user is routed to the target resolver through one or more relays, i.e., multiple hops.
Then, the first key feature of the \schemeabbrv is to enforce each user to choose his dedicated relay as his \emph{next-hop}.
Otherwise, a user himself becomes a trusted relay, or the user chooses a fully-trusted shared relay like the one deployed by his organization, unlike public relays.
The employment of a dedicated and trusted relay as the next-hop simply precludes the collusion between the target resolver and a node that directly has the knowledge of the user's IP address.
Note that this concept of the dedicated next-hop is much like the \emph{entry guard node} in Tor \cite{tor}.

\vspace{1ex}
\noindent\textsc{\textbf{2) Mutually sharing user's dedicated relays to mix query messages:}}
As we mentioned in Section~\ref{sect:formulation}, the identity of the user is uniquely associated with his dedicated relay in the use of a single relay, i.e., ODoH and ADNSCrypt, since the dedicated relay of a user is uniquely coupled with the user himself.
Even in the case of multiple relays, this occurs as well when a hop after the dedicated next-hop of the user has colluded with the target resolver.
Hence the second key feature designed in \schemeabbrv is to enforce the dedicated next-hop of a user to accept incoming messages from other relays or resolvers and forward them in addition to DNS messages from/to the user.
Namely, users mutually shares and leverages their dedicated relays each other as hops after their next-hop.
This allows the next-hop of a user to mix query messages from its user with ones from other relays.
Thus this feature makes it difficult for other relays receiving messages from the next-hop to correctly identify the user's messages.
Namely in our concept of \schemeabbrv, when one shares its resource, i.e., allowing one's dedicated relay to forward messages of others, the privacy for one's identity is protected.

\vspace{1ex}
\noindent\textsc{\textbf{3) Randomization of subsequent hops by the user:}}
To fully leverage the first and second features to hide the user's identity, the choice of relays after the next-hop is quite important.
Consider a case where all relays receiving/sending messages from/to the dedicated relay have unfortunately colluded with the target resolver.
Then the user's message could be easily identified at the target resolver.
This might be unrealistic in general but could be realistic if the target resolver selectively increases colluded nodes when the route is always fixed.
Thus to disallow such a case and minimize the identity leakage even if we meet the case, we introduce the third key feature; the user randomizes choices of hops after his dedicated relay.
This means that the number of subsequent hops, the selection of hops, and the order of hops are all randomized.
That is, the \schemeabbrv allows messages to be conveyed via one or more relays or \emph{directly} to the target resolver after the dedicated next-hop.
Note that in our motivation and in the nature of DNS, only the origin of a query message must be anonymized to its target resolver and its colluded relays, but \emph{the target resolver itself is NOT required to be hidden to relays.}
Hence in choices of subsequent relays, the direct forwarding from the dedicated relay to the resolver is also allowed like ODoH \cite{Singanamalla2020} and unlike Tor \cite{tor}.

\vspace{1ex}
From the mechanical point of view in \schemeabbrv, every DNS query message conveyed on a path includes its forwarding information only about subsequent relays, and the origin of paths is hidden from the target resolver.
For instance in Figure~\ref{fig:modns_concept1}, a query message $Q_3$ on the path between $R_A$ and $R_C$ includes forwarding information that will be forwarded to $X$ after $R_C$, and does not include previous paths, i.e, $A$ to $R_A$.
That is, each relay simply strips previous hops information of an incoming query message.
Note that as with ODoH and ADNSCrypt, DNS response messages from the target resolver include no information of their returning path and that they are correctly returned to the user by trailing L4 connections\footnote{UDP/TCP connections are managed at each node.} between adjacent nodes.


In the next section, we introduce a proof-of-concept implementation of \schemeabbrv with these key features.

\section{Proof-of-concept implementation of \schemeabbrv}\label{sect:poc}
As a proof-of-concept (PoC) of \schemeabbrv, we implemented a client module translating Do53 to \schemeabbrv and a relay module by forking and modifying open-source softwares, dnscrypt-proxy \cite{dnscrypt-proxy} and encrypted-dns-server \cite{encrypted-dns-server}, that supports ADNSCrypt \cite{Denis2021,Denis2020}.
Namely, our PoC implementation is based not on ODoH but fully on ADNSCrypt, and the target resolver should be the one supporting DNSCrypt protocol.
The client module and relay module are available on GitHub at \cite{modns-proxy} and \cite{modns-server}, respectively.
Our publicly available relays are also listed at \cite{modns-resolvers-relays}.

\begin{remark}
At this point (March 1, 2021), there are no public client translating Do53 to ODoH as far as we know \footnote{On March, 30, 2021, dnscrypt-proxy \cite{dnscrypt-proxy} just supported ODoH.}.
Hence, we implemented our proof-of-concept of \schemeabbrv based on the ADNSCrypt protocol.
\end{remark}

To help gain the reader's understandings, we begin with a brief explanation of the ADNSCrypt protocol.
Then, we explain how ADNSCrypt protocol has been modified for \schemeabbrv.
From now on, we call our PoC protocol and implementation of \schemeabbrv as the \emph{PoC \schemeabbrv}.

\subsection{ADNSCrypt protocol}\label{sect:adnscrypt_protocol}
In the ADNSCrypt protocol \cite{Denis2021,Denis2020}, a user and a target resolver exchange all encrypted DNS messages, i.e., queries and responses, via a single relay specified by the user.
The message encryption is done in the end-to-end manner of DNSCrypt version 2 protocol \cite{Denis2020a} between the user and resolver, and the relay only passively forwards messages upstream and downstream.
Thanks to the end-to-end encryption between the user and the target resolver, the relay learns nothing about plaintext DNS queries and responses.
Also thanks to the relay, the specified target resolver learns nothing about the user's IP address unless it colludes with the relay.

To make the relay correctly forward each query message, the user constructs the message in the following form\footnote{Values in target address and port are just examples.}:
\begin{align}
\text{ADNSCrypt query} &:= |\text{ADNSCrypt header}|
\underbrace{|\text{ADNSCrypt payload}|}_\text{= DNSCrypt query},\nonumber\\
\text{ADNSCrypt header}
 &:=
\underbrace{|\mathtt{0xFFFFFFFF\ 0xFFFFFFFF\ 0x0000}|}_\text{Anonymized query magic}
\underbrace{|\texttt{192.168.1.1}|}_\text{Target address}
\underbrace{|\texttt{8443}|}_\text{Target port}, \label{eq:adnscrypt_query}
\end{align}
where the anonymized query magic is a special constant, and the target address and port are variables of those of the target resolver of DNSCrypt v2.
Note that the ADNSCrypt payload part in the above is exactly the encrypted query in the \emph{non-anonymized} DNSCrypt v2,
and hence a query message in ADNSCrypt can be viewed as an encrypted query in DNSCrypt v2 with an extra header indicating the target resolver address and port.
The user simply dispatches this query message over UDP or TCP to its specified relay instead of the target resolver.

When a relay receives a message starting with the anonymized query magic, it peels off the message header part, and forwards only the DNSCrypt query part to the target resolver specified in the peeled header.
Thus, from the viewpoint of the target resolver, it communicates with the relay in the exactly same fashion as the non-anonymized DNSCrypt v2.
The response to the query is structured exactly in the form of the non-anonymized DNSCrypt response, and it is just inversely forwarded from the resolver to the user via the relay with no modification on the path.

\subsection{Specification and implementation of the PoC \schemeabbrv}\label{sect:modns_spec}
To realize the concept of \schemeabbrv, our PoC \schemeabbrv rebuilds the construction rule and format of the query message and modifies the relaying protocol of ADNSCrypt.
In the following, we shall give the detailed explanation of these modifications.

\subsubsection{Query message format and protocol}\label{sect:modns_protocol}
In \schemeabbrv, the target resolver is the DNSCrypt v2 resolver like ADNSCrypt, but note that the message format and relaying protocol are modified and extended from ADNSCrypt.
When a user generates a query message of \schemeabbrv, the message itself has to describe the path on which it will follow to the target resolver.
To this end, we adopt a Type-Length-Value (TLV)-like format for query messages, given as follows.
\begin{align}
\text{\schemeabbrv query} &:= |\text{\schemeabbrv header}|
\underbrace{|\text{\schemeabbrv payload}|}_\text{= DNSCrypt query},\nonumber\\
\text{\schemeabbrv header} &:=
\underbrace{|\mathtt{0xFFFFFFFF\ 0xFFFFFFEE\ 0x0000}|}_\text{\schemeabbrv query magic}
\underbrace{|\mathtt{n}|}_\text{Number of subsequent nodes}
\nonumber\\
&\qquad \underbrace{|\texttt{192.168.1.1}||\texttt{8443}|}_\text{Node $1$ address and port}
\underbrace{|\texttt{192.168.2.4}||\texttt{8443}|}_\text{Node $2$ address and port} \dots
\nonumber\\
&\qquad \dots
\underbrace{|\texttt{192.168.128.32}||\texttt{8443}|}_\text{Node $n$ (target resolver) address and port},
 \label{eq:poc_modns_query}
\end{align}
where the \schemeabbrv query magic is a constant indicating the \schemeabbrv query message much like anonymized query magic in ADNSCrypt.
Supposed that for the query, the user chooses $n$ $(n > 0)$ relays including his dedicated next-hop, and that the \schemeabbrv query message will be forwarded to the target resolver via the selected $n$ intermediate relays.
Then, we see that it has $n$ nodes, i.e., total $n-1$ relays and the resolver, after the next-hop relay on the path.
The message simply describes this path information by the number of such nodes, $n$, and an ordered array of $n$ tuples of IP address and port regardless of relays or a target resolver, as given above.
The query message is routed in order from the beginning node of the array after the user's next-hop, where the end of the array represents the target resolver.
In the above example, after the message is forwarded to the user's next-hop, it is routed towards Node $1$ (192.168.1.1:8443) from the next-hop, and the final destination is Node $n$ (192.168.128.32:8443), i.e., the target resolver.

We note here that the user randomly chooses intermediate relays, and that the number $n$ is also chosen at random as explained in Section~\ref{sect:overview}.
In our PoC implementation, this can be done by specifying the list of available relays and maximum/minimum possible $n$, which will be described later.

When an intermediate relay including the next-hop receives a message starting with the \schemeabbrv query magic, it decrements the number of subsequent nodes and peels off the first element of the array.
For example, when the user's dedicated next-hop received the above message, it first learns from the header that the message has to be routed to Node 1.
Then, it simply decrements the value $n$ to $n-1$, removes the first element, [192.168.1.1:8443], of the array, and sends the restructured message to Node 1.
Note that regardless of the next-hop or not, every relay works in this manner for the received \schemeabbrv query message if $n > 1$.
If $n=1$, the array includes only the information of the target resolver supporting DNSCrypt.
Hence the relay that received such the message strips off all the header and simply dispatch only the part of \schemeabbrv payload, i.e., DNSCrypt query, to the target resolver.
As with the ADNSCrypt, the response to the query is given as that of the vanilla DNSCrypt and inversely forwarded from the resolver to the user.

\subsubsection{Client implementation}\label{sect:modns_client}
Our client implementation \cite{modns-proxy} of the PoC \schemeabbrv protocol given in Section~\ref{sect:modns_protocol} is based on \emph{dnscrypt-proxy} \cite{dnscrypt-proxy}.
The dnscrypt-proxy converts plaintext queries of Do53 to encrypted ones of DoH/DNSCrypt and vice versa for responses.
Since the dnscrypt-proxy also supports ADNSCrypt, i.e., it can convert Do53 queries to ones given as \eqref{eq:adnscrypt_query}, we simply extended this function of dnscrypt-proxy to implement a client of the PoC \schemeabbrv.
In our client implementation, PoC \schemeabbrv queries given as \eqref{eq:poc_modns_query} are generated from Do53 queries for some given parameters such as lists for candidates for the user's next-hop, i.e., dedicated relays, and available relays after the next-hop.
Specifically, the following parameters are newly introduced in addition to the existing parameters of dnscrypt-proxy:

\vspace{1ex}
\noindent\textsc{\textbf{[List of available relays with `next-hop' flag]}}:
Like the vanilla dnscrypt-proxy, available relays supporting the PoC \schemeabbrv are given as a static list or ones hosted online, e.g., \cite{modns-resolvers-relays}.
As metadata, every available relay in a list can have a flag indicating a candidate of the next-hop.
Namely, relays dedicated to the user are explicitly specified by this flag in our client implementation.

\vspace{1ex}
\noindent\textsc{\textbf{[Maximum and minimum number of relays]}}:
For each query message, the number of relays after the next-hop, i.e., $n-1$ in \eqref{eq:poc_modns_query}, are randomly fixed within the range specified by the user.
The range is simply given by parameters of the maximum and minimum allowed numbers of relays.

\vspace{1ex}
When a Do53 query is given from the user, our client implementation first chooses a \emph{flagged} relay candidate as the next-hop from given lists and simultaneously generates a DNSCrypt v2 query from the plaintext Do53 query.
Then, it constructs a PoC \schemeabbrv query as \eqref{eq:poc_modns_query} by selecting a target DNSCrypt resolver and randomly choosing relays from the rest of the listed relays regardless of the flag.
Finally, the generated query is dispatched to the next-hop flagged relay chosen at first.
In our client implementation, the PoC \schemeabbrv query is generated in such a way that any loop path, i.e., duplicated selection of relays, is avoided.
Also note that in our client implementation, any DNSCrypt resolver can be chosen as the target resolver in the same manner as ADNSCrypt.

The detailed configuration is given in the GitHub repository \cite{modns-proxy}.

\subsubsection{Relay implementation}\label{sect:modns_relay}
The second piece of our PoC \schemeabbrv is the relay implementation \cite{modns-server} supporting the \schemeabbrv protocol of Section~\ref{sect:modns_protocol}.
Since \emph{encrypted-dns-server} \cite{encrypted-dns-server} works not only as a DNSCrypt resolver but also as a relay of ADNSCrypt, we simply extended the relay function of the software so as to handle PoC \schemeabbrv queries defined in \eqref{eq:poc_modns_query} in addition to standard ADNSCrypt queries in \eqref{eq:adnscrypt_query}.
The extended part for the PoC \schemeabbrv has the following function to adequately serve PoC \schemeabbrv queries according to the protocol in Section~\ref{sect:modns_protocol}.

\vspace{1ex}
\noindent\textsc{\textbf{[Overload and loop avoidance]}}
At a relay, each incoming query message explicitly indicates subsequent relays where the message will visit after the relay.
If the number of such relays is unnecessarily large, relays might be severely overloaded, and hence such cases should be suppressed.
Towards this end, our relay implementation checks the number of such subsequent relays indicated in each incoming query and drops the one having more hops than the predefined threshold.
It also checks duplicated relays, i.e., loop, in the path of relays indicated in each incoming query, and drops it as well when a loop is detected.\footnote{Loop avoidance is not necessary as long as only our client implementation in Section~\ref{sect:modns_relay} is used.}

\vspace{1ex}

We note that in our relay implementation, the backward path from the target resolver is managed by the socket connection of UDP at each relay as well as the relay function in the original encrypted-dns-server.
For the detailed configuration of our relay implementation, please refer to the GitHub repository \cite{modns-server}.

\subsection{Publicly available relays of PoC \schemeabbrv}\label{sect:deployment}
In a manner the same as relays of ADNSCrypt \cite{dnscrypt-resolvers-relays}, instances of our relay implementation have been deployed and are publicly available on the Internet \cite{modns-resolvers-relays}.
Thus, as we explained in Section~\ref{sect:overview}, a user can choose them as relays after his dedicated next-hop, or simply use them even as the next-hop if he could trust the relay operator (us).
We also deployed public resolvers supporting DNSCrypt v2 as shown in \cite{modns-resolvers-targets}, which can be configured as target resolvers of PoC \schemeabbrv in addition to existing ones, e.g., \cite{dnscrypt-resolvers-targets}.
Our relay and resolver instances are being operated on VPS's (virtual private servers) in Tokyo and Singapore.
In the following section, we shall give a small performance evaluation using these instances.

\section{Evaluation and discussion}\label{sect:discussion}

This section first gives an analysis of the security and privacy of the \schemeabbrv, and discusses its limitations.
We then introduce a preliminary evaluation of the PoC \schemeabbrv given in Section~\ref{sect:poc} in terms of the round trip time (RTT) by a small-scale deployment on the Internet.

\subsection{Security, privacy and limitation}\label{sect:security_evaluation}
Our concept \schemeabbrv can be viewed as an extension of single relay-based schemes, i.e., ADNSCrypt and ODoH, to multiple relay-based ones.
Namely, we see that by omitting the assumption of leveraging the user's dedicated next-hop relays, the \schemeabbrv guarantees at least the same security and privacy as ADNSCrypt and ODoH.

The \schemeabbrv additionally introduces settings that dedicated next-hops are given for users and that they serve incoming queries from external users as relays after next-hops.
Here we consider a case where there are two different users with dedicated relays as employed as their next-hops, and the relays are connected with other parties' ones.
Then from the viewpoint of one of the two next-hops, incoming messages are randomly mixed ones dispatched from the other relay, i.e., the other user, and third party users.
We thus see that each of next-hops cannot uniquely identify messages of the other user among incoming messages unless it have colluded with all of third party users/relays.
We thus deduce that even if a subset of relays colludes with target resolvers, the target resolver cannot identify the origin of incoming queries since no colluded relay always identifies the origin of queries relayed.

Note that the collusion resistance of \schemeabbrv is NOT universal, i.e., not always guaranteed in any environment, since the location of colluded relays is unknown to users.
In particular, for a fixed path from a user to a target resolver, the user's identity is leaked to the target resolver when all of the relays connected to the user's next-hop have colluded with a target resolver.
Thus by the key feature 3 of randomization in Section~\ref{sect:overview}, the collusion resistance of \schemeabbrv is ensured in a \emph{probabilistic} manner that the user cannot be identified by the target resolver unless selected paths meet such the condition.
Consider that a number of entities make relays publicly available in the system. We claim that in this case, the leakage of the user identity quite rarely occurs if relays after the next-hop are randomly chosen from ones operated by mutually different entities, like the limitation of Tor \cite{tor}.

As a conclusion of this subsection, we mention the security of our PoC implementation.
Although we have assumed no wiretappers on transit channels in Section~\ref{sect:formulation}, an extra layer of security might be required.
For instance, there may be a case where a certain monitoring authority employs a middlebox that drops any transit messages conveyed to specific relays/resolvers.
Recall that our PoC \schemeabbrv is based on ADNSCrypt and all the connections among relays are established over UDP, i.e., no channel encryption like TLS.
Considering channel wiretappers, the target resolver and path of relays are explicitly revealed from an observed query message while the content of the query itself is encrypted.
Therefore, if channel wiretappers are considered, we should employ the channel encryption for relaying, and ODoH-based \schemeabbrv is such a candidate to enhance the security by HTTPS connections for relaying.

\subsection{A preliminary performance evaluation}\label{sect:performance_evaluation}

We evaluate the performance of the PoC \schemeabbrv from the viewpoint of round-trip-time (RTT), i.e., the elapsed time to receive a response after a query issuance under several relay settings.

\begin{table}
\centering
\begin{tabular}{|l||l|l|}
\hline
\multirow{4}{*}{\textbf{Node setting}}
& PoC \schemeabbrv User \cite{modns-proxy} & ConoHa VPS Tokyo \\
\cline{2-3}
& Target resolver \cite{encrypted-dns-server} & ConoHa VPS Tokyo \\
\cline{2-3}
& \multirow{2}{*}{PoC \schemeabbrv Relays \cite{modns-server}} & 3 nodes: ConoHa VPS Tokyo\\
& & 2 nodes: ConoHa VPS Singapore\\
\hline
\textbf{\# of measurements} & \multicolumn{2}{l|}{$10,000$ times} \\
\hline
\multirow{4}{*}{\textbf{Choices of relays}}
 & \multicolumn{2}{l|}{1) Random Tokyo $t$ relays ($t=0,\dots,3$)}\\
 & \multicolumn{2}{l|}{2) Random Singapore $s$ relays ($s=1,2$)}\\
 & \multicolumn{2}{l|}{3) Random Tokyo $1$ and then Singapore $1$ relays}\\
 & \multicolumn{2}{l|}{4) Random Singapore $1$ and then Tokyo $1$ relays}\\
\hline
\end{tabular}
\caption{Environments of our experimentation.}\label{table:exp_node_location}
\end{table}
\begin{table}
\centering
\begin{tabular}{|l|l|l|}
\hline
& \textbf{Relays} & \textbf{Average RTT}\\
\hline
\hline
\multirow{4}{*}{1)}
 & Direct ($0$ relays = DNSCrypt) & $126.3$ ms\\
\cline{2-3}
 & Tokyo $1$ relay (= ADNSCrypt) & $137.0$ ms\\
\cline{2-3}
 & Tokyo $2$ relays & $137.5$ ms\\
\cline{2-3}
 & Tokyo $3$ relays & $136.1$ ms\\
\hline
\multirow{2}{*}{2)}
 & Singapore $1$ relay (= ADNSCrypt) & $289.8$ ms\\
\cline{2-3}
 & Singapore $2$ relays & $293.9$ ms\\
\hline
3) & Singapore $1$ relay then Tokyo $1$ relay & $288.3$ ms\\
\hline
4) & Tokyo $1$ relay then Singapore $1$ relay & $215.8$ ms\\
\hline
\end{tabular}
\caption{Results of the experimental evaluation: Average RTTs for DNS query/response under several conditions of intermediate relays between the user in Tokyo and the resolver in Tokyo.}\label{table:exp_result}
\end{table}

\subsubsection{Experimental environment}
Table~\ref{table:exp_node_location} summarizes the environment of our experimentation, where all the nodes, i.e., a dnscrypt-proxy-based user, relays, and a target resolver, are deployed on virtual private servers (VPS's) \cite{conoha} at Tokyo or Singapore.
Under this setting, we measured the RTT from the user's query issuance to response retrieval with several choices of relays.
In each measurement, the user generates a query for a domain with a random subdomain based on UUIDv4, e.g.,
\begin{align*}
\texttt{41587c8e-6adf-4d2e-af63-c35e2bd19c7c.example.com},
\end{align*}
to disable caching at the target resolver and measure the RTT.
We measured the average RTT of $10$k measurements for each setting of relays, where we fixed the number of relays chosen and the user randomly selected relays of the fixed number from the pool of candidate relays.
For the choice of relays, we executed four types of experimentation;
1) With relays located in Tokyo, 2) With relays located in Singapore, and 3/4) With two relays of one located in Tokyo and the other located in Singapore.

\subsubsection{Experimental result}
The average RTTs of $10$k measurements are summarized in Table~\ref{table:exp_result} for each type of experimentation.
For 1) measurements with relays in Tokyo, we see that average RTTs with more relays are comparable to and not degraded from the case of direct connection (0 relays), i.e., the vanilla DNSCrypt v2, and that of $1$ relay, i.e., ADNSCrypt.
Also for 2) measurements with relays at Singapore, the result with $1$ relay is exactly similar to the case of $2$ relays.
For 3) and 4) measurements with a combination of relays from Tokyo and Singapore, their average RTTs are larger than that of 1) and comparable to 2).

\subsubsection{Consideration from the experimental result}
As easily expected, we can view that from the experimental results, the performance in terms of RTT simply and mostly depends on geographic locations of intermediate relays in our PoC implementation of \schemeabbrv.
We also see that from the result for the case 1) and 2), the relaying operation described in Section~\ref{sect:modns_relay} does NOT become a considerable overhead.

From the above observation, a user should choose relays after his dedicated next-hop from a set of its neighbor nodes to gain better performance.
However, such a restriction of candidates of subsequent relays could cause privacy degradation.
This is because as follows.
Recall that subsequent relays are randomly chosen from such a set of candidate nodes and that the privacy of \schemeabbrv is guaranteed under the assumption that a tiny subset of the set is colluded, as explained in Section~\ref{sect:security_evaluation}.
This means that to realize better privacy, the set of subsequent relay candidates should be large.
Thus as a simple operational guideline, we suggest that for the maximum capable RTT, the set of subsequent relay candidates should be as large as possible.

As the conclusion of this section, we claim that in general, the RTT less than a few hundred milliseconds, e.g., $\le 500$ ms, does not degrade the user experiences.
We thus conclude that the PoC \schemeabbrv achieves a moderate performance the at least in our small experimentation.

\subsection{Expected deployment scenarios}\label{sect:deployment_scenario}
Considering the deployment of relays, we expect several scenarios for the employment of \schemeabbrv;
For users in an organization, their common next-hop relay is deployed at the organization's network edge, e.g., the gateway. Then it is also made public for external users and used as a relay subsequent to their next-hops.
For individual users in a mobile network, their next-hop relays would be instantiated at their edge-computing nodes.
On the other hand, much like Tor \cite{tor}, some altruistic entities might deploy relays that are transparent and trusted for certain groups of users.

\section{Concluding remarks}\label{sect:conclusion}
By extending the concept of single-relay-based DNS, i.e., ODoH \cite{Singanamalla2020} and ADNSCrypt \cite{Denis2021,Denis2020}, this paper introduced a concept of a multiple-relay-based DNS for user's anonymity in DNS queries, called the mutualized oblivious DNS (\schemeabbrv).
In the \schemeabbrv, the user just sets the dedicated one as his \emph{next-hop}, i.e., first relay, conveying his queries to the resolver and randomly chooses its subsequent relays shared from other entities.
Under this setting of sharing a small resource, the user's identity is concealed to a target resolver even if a certain (unknown) subset of relays have colluded with the target resolver.
The concept of the next-hop in \schemeabbrv could be viewed as an approach similar to the entry guard nodes in Tor \cite{tor}, but the design of \schemeabbrv is highly specialized and simplified to the DNS-specific nature.
Moreover, we have introduced a PoC implementation of \schemeabbrv based on ADNSCrypt, and evaluated its performance in small experimentation.
In the experimentation, it was demonstrated that our PoC implementation achieves a moderate performance in terms of the RTT, which is comparable to existing single-relay-based schemes.

Our study on \schemeabbrv is still preliminary and there exist many rooms to be considered as future works.
For instance, since only a small-scale measurement of RTT has been done in this paper, we should conduct a larger scale measurement of the performance by deploying numbers of relays at geographically distributed regions.
Another possible avenue is to enhance our PoC implementation based on ADNSCrypt to the one based on ODoH, and to measure its performance with the utilization of CDNs.


\section*{Acknowledgment}
This work was supported in part JSPS KAKENHI Grant Number JP20K23329, and KDDI Research collaborative research project.
The authors would like to thank Prof. Toshiaki Tanaka at University of Hyogo for his valuable comments.

\bibliographystyle{IEEEtranS}
\bibliography{./references.bib}

\end{document}

%% file: symdef.tex
\theoremstyle{definition}
\newtheorem{theorem}{Theorem}

\newtheorem{remark}[theorem]{Remark}

\def\ie{{i.\@e.\@,~}}
\def\eg{{e.\@g.\@,~}}